# Radio-Frequency Polymer Field-Effect Transistors Characterized by S-Parameters

M. Giorgio and M. Caironi

*Abstract*—Thanks to recent progress in terms of materials properties, polymer field-effect transistors (FETs) operating in the MHz range can be achieved. However, further development towards challenging frequency ranges, for a field accustomed to slow electronic devices, has to be addressed with suitable device design and measurements methodologies. In this work, we report n-type FETs based on a solution-processed polymer semiconductor where the critical features have been realized by a large-area compatible direct-writing technique, allowing to obtain a maximum frequency of transition of 19 MHz, as measured by means of Scattering Parameters (S-Parameters). This is the first report of solution-processed organic FETs characterized with S-Parameters.

## I. Introduction

ORGANIC electronics underwent an impressive development and thanks to its peculiar properties, e.g. high degree of tunability of electronic properties and thin molecular films deposition with scalable processes at low temperature on a wide variety of substrates, represent one of the most promising candidates in the path to ubiquitous, flexible and lightweight low-cost electronics.
To date, organic light emitting diodes (OLEDs)[1] have reached commercial maturity, and significant progress has been achieved for organic solar cells[2], photodetectors[3], biosensors[4] and circuits based on organic transistors[5]. In the case of field effect transistors (FETs), the use of polymer semiconductors has the advantage of achieving superior mechanical properties and easily enabling large-area fabrication techniques, such as printing. Excellent DC performances have been recently achieved owing to an improvement in the understanding of device physics[6], and steady increase in charge carrier mobility[7] and charge carrier injection[8]. Yet, progresses in DC performances are not usually matched by corresponding improvements in AC properties, since typical FET structures devised to test new polymer semiconductors are not optimized for frequency operation. It is instead highly demanded to enhance polymer FET maximum operational frequency,

Manuscript received Month xx, 2019; accepted Month xx, 201x. Date of publication Month xx, 201x; date of current version Month xx, 201x. This work was supported by the European Research Council, European Union's Horizon 2020 Research and Innovation Programme HEROIC under Grant 638059. The review of this paper was arranged by Editor Name. M. Giorgio and M. Caironi are with the Center for Nano Science and Technology @PoliMi, Istituto Italiano di Tecnologia, 20133 Milano, Italy (e-mail: mario.caironi@iit.it). M. Giorgio is with Dipartimento di Elettronica, Informazione e Bioingegneria, Politecnico di Milano, Piazza Leonardo da Vinci 32, Milano, Italy.

as this would enable a much broader application of this technology. For example, higher operational frequencies are crucial in drivers for high resolution flexible displays[9], radio frequency identification (RFID) systems for smart items identification[10] or wireless communication[11].

A widely adopted figure of merit to assess the maximum operational frequency of transistors is the frequency of transition ($f_T = \frac{g_m}{2\pi C_{TOT}}$, where $g_m$ is the transconductance and $C_{TOT}$ is the total gate capacitance), or unity gain frequency[12], defined as the frequency at which the ratio of the absolute values of the small-signal ac gate current and the small-signal ac drain current equals one. Even if polymer FETs with MHz-range $f_T$ have been achieved, so far their measurement has been complicated by parasitism introduced by the characterization setup itself. In the framework of organic FETs reports, the latter typically becomes dominant already around 1 MHz, and can make the $f_T$ parameter extraction unfeasible already in the tens of MHz range. The reason lies in the commonly adopted characterization techniques that can be referred as *direct measurements*, consisting in directly measuring, typically with low frequency equipment, the AC gate current $i_g$ and the AC drain current $i_d$ as a function of a modulating gate voltage, and then by evaluating the frequency of the crossing point between the two curves. As a consequence, $f_T$ for organic FETs above 10 MHz has been mainly determined so far thanks to an extrapolation from the data obtained at low frequency.[13,14,15] In a single case, for a FET based on evaporated small molecules on electrodes patterned by conventional photolithographic techniques, an $f_T$ of 20 MHz has been measured without extrapolation, thanks to the adoption of inductive probes[16]. For polymer FETs instead, an $f_T$ above 10 MHz was achieved thanks to extrapolation in two cases, in which the device was fabricated by combining direct-writing and printing techniques[17,18].

Notably, according to recent works, $f_T$ in the GHz range is in principle achievable with organic transistors if a series of requirements are met[19,20,21]. Therefore it is not possible to rely further on direct measurement methods for characterizing the upcoming organic high-frequency devices. For this reason, two-port Scattering Parameters (S-Parameters)[22], adimensional quantities which relate the AC currents and voltages between the drain and the gate contacts, become essential also for organic FETs as they readily allow to widen the measurement bandwidth up to the GHz range. Starting from them, a series of parameters can be mathematically computed, such as admittance parameters (Y-Parameters)[22], which are useful to extract physical device parameters, e.g. device capacitances.

Accordingly, organic FETs, which are typically far from being standardized, have to be fabricated compatibly with S-Parameters characterization tools. So far, S-parameters were adopted only by one group for measuring a maximum $f_T$ of 6.7 MHz in organic FETs based on an evaporated small-molecule semiconductor, and realized by using silicon stencil masks to define critical lateral features in the sub-micron range[23,24,25]. At present, no reports describe the use of S-Parameters to measure $f_T$ in the case of FETs based on solution processed organic semiconductors.

In this work, we adopted S-parameters to characterize solution-processed polymer FETs where critical features have been realized by a direct-writing, large-area compatible technique. We fabricated and characterized polymer FETs having channel lengths varying from 3.7 µm down to 1.2 µm and the gate to source and drain overlap lengths ranging from 2.6 µm to 2.3 µm, assessing the scaling of the corresponding $f_T$ at different bias points. For the shortest channel length transistor we obtained a maximum $f_T$ of 19 MHz at the relatively low applied gate-source voltage $V_{GS}$ of 12 V.

## II. OFETS FABRICATION PROCESS

A schematic representation of the FET fabrication process flow is illustrated in Figure 1. A 30 nm thick gold pattern, which only serves as interconnection between the probe needles and the actual transistor, was fabricated by standard photolithography on a glass substrate. Then, a polymer transistor was fabricated entirely by means of mask-less processes. In order to achieve high-frequency operation, we adopted a direct-writing method for the definition of the FET critical features. In particular we made use of a femtosecond laser sintering technique for the fabrication of high resolution source and drain metal electrodes, crucial for both the realization of micrometer size channel lengths and the reduction of overlap parasitic capacitances[17]. Such process starts with a uniform deposition of a silver nanoparticle ink film by spin-coating (Fig. 1a). Then, a femtosecond pulsed laser beam ($\lambda$ = 1030 nm, 67 MHz repetition rate) locally heats the nanoparticle film and patterns the desired geometry of conductive features (Fig. 1b). The unprocessed areas of the film are then washed away using o-xylene, and only the conductive high-resolution patterns are left on the substrate (Fig. 1c).

After the contacts fabrication, the substrate was cleaned by argon plasma and source and drain contacts were modified by growing a 4-(dimethylamino)benzenethiol (DABT) self-assembled monolayer, a process that reduces the contact work function and consequently the energetic barrier for electrons injection into the semiconductor[26]. Afterwards the very well-known semiconducting co-polymer poly[N,N'-bis(2-octyldodecyl)-naphthalene-1,4,5,8-bis(dicarboximide)-2,6-diyl]-alt-5,5′-(2,2′-bithiophene), P(NDI2OD-T2), was deposited by off-center spin-coating in order to achieve uniaxial alignment of polymer backbones, thus improving the charge-carriers transport properties along the alignment direction[27]. Before depositing a 90 nm thick layer of parylene-C, a 20 nm layer of PMMA was deposited by spin-coating to guarantee optimal interfacial properties for an efficient charge transport at the accumulated channel-dielectric interface (Fig. 1d).

To complete the device, via-holes to the gate pad were fabricated by laser ablation (Fig. 1e) and a silver top gate contact was inkjet printed on the channel area using a Fujifilm Dimatix printer and an Ag nanoparticles based ink. The silver layer also connects the gate electrode to the gate interconnection pad (Fig. 1f). Finally, devices were encapsulated by depositing a 500 nm thick parylene capping layer, on top of which a bicomponent epoxy resin (purchased from Robnor Resinlab) was deposited.

## III. DC CHARACTERIZATION

We fabricated a set of transistors having channel lengths ($L$) of 3.7 µm, 3.1 µm, 1.8 µm and 1.2 µm and channel width ($W$) of 1.2 mm. Transfer curves in the linear and saturation regime of operation are reported in Figure 2a. Devices exhibit good n-type behavior, showing ON/OFF current ratios of at least 5 orders of magnitude despite the short channels. The ON current at $V_{GS}$ = 12 V perfectly scales with $L$ in the linear regime of operation ($V_{DS}$ = 5 V, Fig. 2a), while the scaling is slightly superlinear in the saturation regime of operation ($V_{DS}$ = 15 V, Fig. 2b). This can be also visualized by looking at the channel width normalized transconductance in saturation ($g_m/W$, Fig. 2c), which increases superlinearly with respect to $V_{GS}$, reaching at 12 V the value of 0.16 µS/µm for $L$ = 1.2 µm and 0.03 µS/µm for $L$ = 3.7 µm. The superlinear behavior in saturation becomes more pronounced as the channel length is shortened, as a result of the effect of the higher lateral field on injection and/or transport[28,29,30]. Correspondingly, the apparent mobility ($\mu_{app}$), extracted by computing the derivative of the transfer curve with respect to $V_{GS}$ and assuming the validity of the gradual channel approximation, is substantially constant with respect to $L$ in the linear regime, whilst it increases with decreasing $L$ in the saturation regime. The maximum linear and saturation $\mu_{app}$ are 0.35 cm$^2$/Vs and 1 cm$^2$/Vs respectively, and are obtained for the shortest channel length transistor ($L$ = 1.2 µm). The corresponding effective mobility in saturation, i.e. the apparent mobility corrected by the reliability factor[31], is 0.4 cm$^2$/Vs.

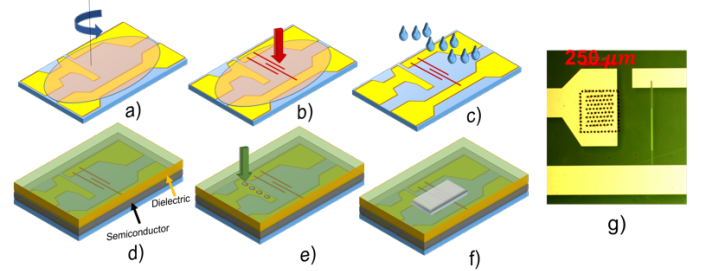

Fig. 1. Sketch of the FETs fabrication process. a) Spin-coating of the nanoparticle ink. b) Laser sintering of source and drain lines. c) Washing off of the unsintered ink. d) Deposition of semiconductor and dielectric layers. e) Fabrication of via-holes by laser. f) Inkjet printing of the gate contact. g) Microscope image of the FET prior to gate deposition.

The output characteristics for longer channel lengths are ideal, while an S-shape at low $V_{DS}$ voltage becomes evident as the channel length is decreased, owing to the presence of a contact resistance effect, estimated to be in the order of 1 kΩcm using the TLM method as a first approximation[32]. The output resistance, extracted by taking the inverse of the slope of the output characteristics at high $V_{DS}$, ranges from 35 kΩ ($L$ = 1.2 µm) to 225 kΩ ($L$ = 3.7 µm). The intrinsic gain, calculated by multiplying the output resistance by the transconductance, ranges from 6.65 ($L$ = 3.7 µm) to 7.9 ($L$ = 1.2 µm).

## IV. AC CHARACTERIZATION

$f_T$ is the reference figure of merit that we used to assess the maximum operational frequency of our polymer FETs. In order

to derive $f_T$ from S-parameters, the system undergoes first a SOLT (Short-Open-Load-Thru) calibration. By measuring these standards, a 12-terms error correction model[33] is applied to remove non-idealities introduced by the measurement setup itself and to shift the calibration plane at the probe tips.

Then, once the FETs S-Parameters matrix is determined, we compute the hybrid parameter transistor current gain with short-circuited output, $h_{21} = \frac{I_2}{I_1}\Big|_{V_2=0}$, where $I_2$ and $I_1$ are the drain and gate small-signal currents, respectively, and $V_2$ is the drain small-signal voltage. To highlight the dependence of $h_{21}$ on physical parameters, its expression can be rewritten by making explicit $I_2$ and $I_1$ as $h_{21} = \frac{g_m - j\omega C_{gd}}{j\omega(C_{gs}+C_{gd})}$, where $C_{gd}$ and $C_{gs}$ are the gate-to-drain and the gate-to-source capacitance, respectively. At

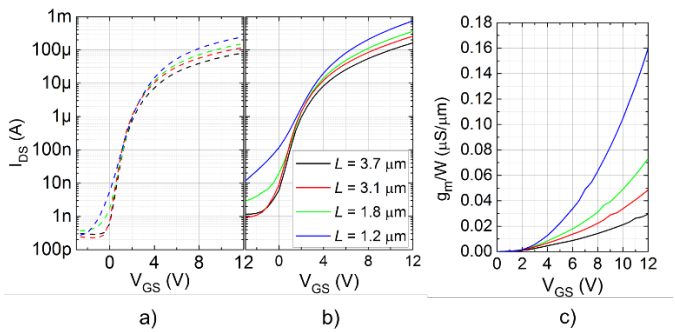

Fig. 2. Transfer characteristics of devices having *L* ranging from 3.7 μm down to 1.2 μm, *W* = 1.2 mm, biased a) in linear regime ($V_{DS}$ = 5 V) and b) in saturation regime ($V_{DS}$ = 15 V). c) $g_m$ as a function of $V_{GS}$, for the different *L* values, extracted from transfer curves in saturation regime, normalized with respect to *W*.

Low frequency $h_{21}$ decreases by 20 dB/decade, while at high frequency it flattens to $C_{gd} / (C_{gs} + C_{gd})$. From its definition, $f_T$ can be identified as the frequency at which $h_{21}$ crosses the 0 dB axis. The measured $h_{21}$ as a function of *L* is shown in Figure 3a.

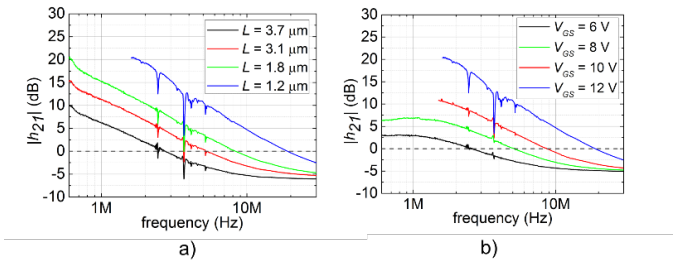

Fig. 3. a) Bode plot of |$h_{21}$| highlighting the transition frequency dependence on a) transistor channel length (with $V_{GS}$ = 12 V and $V_{DS}$ =15 V) and b) bias point of operation for the shortest *channel length* transistor (*L* = 1.2 μm, $V_{DS}$ = 6, 8 ,10, 15 V). Spurious interferences, originated from the DC voltage source adopted, are visible from 2 to 5 MHz.

For the shortest channel length transistor, $h_{21}$ follows the expected trend and at high frequency, $h_{21}$ flattens at -6 dB, in agreement with the capacitances extracted from the admittance parameters. $h_{21}$ crosses the 0 dB axis at $f_T \approx$ 19 MHz. We note the effect of the interconnections series and parallel parasitic impedances, quantified through a two-step de-embedding procedure[34], are negligible in our conditions and for the frequency range of interest.

Such a high $f_T$ is achieved at a relatively low voltage ($V_{GS}$ = 12 V). As a result, this is the fastest organic transistor measured with S-parameters and in general one of the fastest organic transistors if we look at the frequency of transition normalized for the applied gate-source voltage, $f_T/V_{GS}$[18], here reaching 1.6 MHz/V. The measured $g_m$, $C_{TOT}$, gate to source and drain geometric overlap length ($L_{ov}$) and $f_T$ are reported in Table 1 together with $f_T$ expected from first principle calculations, where $g_m$ is extracted from quasi-static transfer curves in Figure 2b, and $C_{TOT}$ is extracted from Y-parameters, being |$Y_{11}$| = $\omega C_{TOT}$. Also $C_{gd}$ can be extracted from Y-parameters according to |$Y_{12}$| = $\omega C_{gd}$, finally obtaining also $C_{gs}$ as $C_{gs} = C_{TOT} - C_{gd}$.

| Channel length | $g_m$ | $C_{TOT}$ | $L_{ov}$ | Calculated $f_T$ | Measured $f_T$ |
|---|---|---|---|---|---|
| *L* = 1.2 μm | 190 μS | 1.27 pF | 2.3 μm | 24 MHz | 19 MHz |
| *L* = 1.8 μm | 90 μS | 1.4 pF | 2.6 μm | 10 MHz | 8.3 MHz |
| *L* = 3.1 μm | 60 μS | 1.47 pF | 2.5 μm | 6.5 MHz | 5.2 MHz |
| *L* = 3.7 μm | 35 μS | 1.46 pF | 2.6 μm | 3.8 MHz | 2.8 MHz |

Table 1. Summary of transistors $g_m$ and $C_{TOT}$ for different *L* ($V_{GS}$ = 12 V and $V_{DS}$ = 15 V) and comparison between measured $f_T$ and computed $f_T$.

In the case of the shortest channel length transistor (*L* = 1.2 μm), we also characterized the dependence of the $h_{21}$ parameter on the FET voltage bias point in saturation, by applying $V_{GS}$ = 6, 8, 10, 12 V and $V_{DS}$ = 6, 8, 10, 15 V, respectively (Fig. 3b). As a result, $f_T$ increases from 2.5 MHz at $V_{GS} = V_{DS}$ = 6 V to 19 MHz at $V_{GS}$ = 12 V and $V_{DS}$ = 15 V. We successfully checked that measured $f_T$ values are consistent with predicted values also in this case by extracting $g_m$ from the static curves of and $C_{TOT}$, constant with bias at 1.27 pF, from the Y-parameters.

## V. Conclusions

In this work, we reported n-type FETs based on a solution processed polymer semiconductor fabricated through a mask-less approach in which critical lateral features, such as source and drain contacts width and channel length, are defined at the microscale by an up-scalable, direct-writing laser-based technique. Such technique allows to increase the FETs maximum operational frequency since it can be used to simultaneously decrease the channel length and the parasitic capacitance due to the gate-source and gate-drain overlap length. For the first time for solution-processed organic semiconductor devices, we resorted to S-Parameters for reliable determination of the frequency of transition without extrapolations, characterizing devices reaching $f_T$ = 19 MHz at a relatively low applied bias of 12 V. As further optimization of devices architecture and reduction of contact resistance, in combination with the use of higher mobility polymers, should allow achieving the hundreds of MHz range, it will be necessary to adopt S-Parameters measurements for the future development of high-frequency polymer electronics.

## Acknowledgement

The authors gratefully acknowledge A. Perinot for helping with the design of the devices and their fabrication, and M. Butti and L. Criante for their support with laser sintering.